\documentclass[a4paper,12pt,openany]{article}
\usepackage[top=20truemm,bottom=20truemm,left=5truemm,right=5truemm]{geometry}
\usepackage[]{hyperref}
\usepackage{amsmath,colortbl,amssymb,amsthm,ascmac,mathrsfs,mathtools,tikz}
\usepackage[]{graphicx}
\newtheorem{example}[]{Example}

\graphicspath{{./figs/}}

\title{Effects of the acceleration on holographic complexity}
\author{Koichi Nagasaki}
\date{\today}

\begin{document}
\begin{center}
	{\LARGE Effects of the acceleration on holographic complexity}\\
\vspace{2cm}
	{\large Koichi Nagasaki}\footnote{koichi.nagasaki24@gmail.com}\\
\vspace{0.5cm}
	{\small Department of Physics, 
	Toho University,\\
	Address: 2 Chome-2-1 Miyama, Funabashi, Chiba 274-8510, Japan}
\end{center}
\vspace{1cm}

\abstract{
In this work we consider a spacial kind of spacetime called AdS accelerating black holes.
This is a kind of black holes which have a stringlike singularity along polar axises attached to the black hole and it accelerates the black hole.
In these kind of spacetimes the growth of Einstein-Hilbert action is independent of the acceleration as found in \cite{Chen:2021qbs}.

By using a string as a probe, we found the effect of the acceleration is captured by the string prove \cite{Nagasaki:2021ldz}. 
Here in this work we consider the case of rotating black holes. 
By the prove string we clearly describe the effect of the acceleration and its relation to the rotation of the black holes.
}


\section{Introduction}
The black holes are main subjects to see the quantum gravitational effects.
The spacetime structure of the black hole should be explained by the quantum gravity theory.
Many theoretical evidences suggests that the black holes are the fastest scrambler \cite{Sekino:2008he}.
So far, we still have the problem, especially about information paradox.
In the context of the development of the inside of the black hole (wormhole), ``complexity" is proposed to be a new physical quantity \cite{Susskind:2014moa}.

Quantum mechanically, complexity is defined as the number of gates needed to form 
the target state from a given initial state.
It has the origin in the quantum information theory and the relation to black holes receives a lot of attention \cite{0034-4885-75-2-022001,2008arXiv0804.3401W,Arora:2009:CCM:1540612, Moore:2011:NC:2086753,Susskind:2014moa, Susskind:2014rva, Caputa:2017yrh, Hashimoto:2018bmb, Susskind:2018fmx, Susskind:2018pmk, Bhattacharyya:2018wym,Chapman:2021jbh}.
Complexity is expected to have a well-known property from thermodynamically view \cite{Brown:2017jil,Carmi:2017jqz,Fan:2019mbp,Bernamonti:2019zyy}.
It also has specific characters in geographical view as a geodesic in circuit space of gates \cite{2005quant.ph2070N,2006Sci311.1133N,2007quant.ph1004D}.

The AdS/CFT correspondence \cite{Maldacena:1997re} predicts that this quantity is calculated by a gravitational quantity.
It is called the ``holographic complexity."
According to the first proposal \cite{Brown:2015bva, Brown:2015lvg}, the two following conjectures are reliable: 
\begin{itemize}
\item {\bf (CV)} 
It is the abbreviation of the ``complexity-voloume" conjecture.
The black hole complexity is proportional to the volume of the Einstein-Rosen bridge (wormhole). 
\begin{equation}
\text{Complexity} = \frac{\text{volume}}{G\ell}
\end{equation}
where $G$ is the gravitational constant and $\ell$ is a particular length scale. 
\item {\bf (CA)} 
It is the abbreviation of the ``complexity-action" conjecture.
This predicts that the black hole complexity is equal to the action defined in the spacetime region called the ``Wheeler-DeWitt patch."
\begin{equation}
\text{Complexity} = \frac{\text{action}}{\pi\hbar}.
\end{equation}
\end{itemize}
The complexity-action (CA) conjecture states that the gravitational counterpart of complexity is the gravitational action \cite{Brown:2015bva, Brown:2015lvg,Alishahiha:2015rta,Chapman:2016hwi,Tao:2017fsy,Jiang:2019fpz,Jiang:2019pgc}.
This statement is checked so far in many works \cite{Alishahiha:2015rta, Chapman:2016hwi, Tao:2017fsy, Jiang:2019fpz, Jiang:2019pgc}.
In this paper we would like to focus on the second conjecture, {\rm i.e.}, we will find a new property of complexity following the calculation of the Wheeler-DeWitt action.

To find more properties of complexity we use the method of probe strings.
This is the method for finding the behavior of quarks moving in a thermal plasma through the drag force by a string \cite{Gubser:2006bz, NataAtmaja:2010hd}.
In the perspective of the AdS/CFT correspondence, a fundamental string or other branes are thought as the non-local objects \cite{Ageev:2014nva,Moosa:2017yvt,Fu:2018kcp,Numasawa:2018grg} {\rm e.g.}, 
the fundamental string is the Wilson loop operator, the D1-branes corresponds to 't Hooft operators, and so on.
In this setting the total action is 
\begin{equation}
S_\text{total} = S_\text{bulk} + S_\text{string}
\end{equation}
where the first term is the gravitational action consisting of the Einsten-Hilbert term, and the boundary terms.  
The perturbation term $S_\text{string}$ which is the Nambu-Goto action measures the deviation from the pure black hole spacetime.

An important reason for using a string to explore the behavior of complexity is the following: 
So far non-trivial features of holographic complexity is found by a local quench \cite{Ageev:2018nye}.
However, it is known that the holographic complexity has non-local property \cite{Fu:2018kcp}.
It should also be consider non-local probes for study the properties of complexity.
It will shed light on new features of holographic complexity and related conjectures. 

In the past work \cite{Nagasaki:2021ldz} we focus on the accelerating black holes. 
This is a special case of more general spacetime described by $\mathbf C$-metric  \cite{Griffiths:2006tk}. 
The property of accelerating black holes is studied \cite{Podolsky:2002nk,Appels:2016uha,Gregory:2017ogk,Anabalon:2018qfv,Appels:2018jcs,Gregory:2019dtq,Zhang:2019vpf,Ball:2020vzo}.
The accelerating properties of such a metric is studied in \cite{Carneiro:2021hgu}.
In the context of the AdS/CFT correspondence a minimal surface is given \cite{Xu:2017nut}. 

In the perspective of the black holes observation, gravitational waves \cite{PhysRevLett.124.251102} caused by such accelerating black hole systems are discovered in the past.
Then $\mathbf C$-metric describing the such an accelerating system is an important subject also from the perspective of observation.

The previous work \cite{Nagasaki:2021ldz} reveals that the string probe detect the acceleration while the Einstein-Hilbert action only depends on the deficit parameter \cite{Chen:2021qbs}. 
Here we would like to study further the case where the black hole has the angular momentum.

\section{Setting}\label{sec:setting}
\subsection{Accelerating black holes}
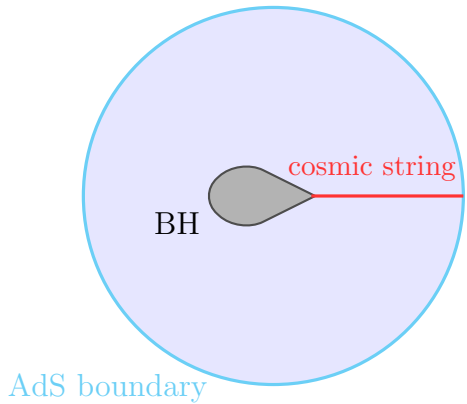
\begin{figure}
\begin{center}
\begin{tikzpicture}
\filldraw[color=cyan!50, fill=blue!10, very thick](0,0) circle (2.5)
 node[xshift = -12ex, yshift = -14ex]{AdS boundary};
\filldraw[fill=black!30, thick, draw=black!70] 
(0.55,0) {[rounded corners, rounded corners=10pt] -- (-0.45,-0.5) -- (-1,0) -- (-0.45,0.5)} -- cycle
  node[xshift = -10ex, yshift = -2ex]{BH};
\draw[color=red!80, very thick](0.5,0) -- (2.5,0) node[pos=0.4, yshift = 2ex]{\small cosmic string};
\end{tikzpicture}
\caption{Structure of the accelerating black hole}\label{fig:accBH_string}
\end{center}
\end{figure}
The acceleration of the black hole is induced by a cosmic string attached on it \cite{Kodama:2008wf}.
The metric is given \cite{Appels:2016uha,Appels:2018jcs} by 
\begin{equation}
ds^2 = \frac1{\Omega^2}\Big[
 -\frac{f(r)}{\Sigma}\Big(dt - a\sin^2\theta\frac{d\phi}{K}\Big)^2
 + \frac{\Sigma}{f(r)}dr^2 + \frac{\Sigma}{g(\theta)}r^2d\theta^2
 + \frac{g(\theta)}{\Sigma}\frac{\sin^2\theta}{r^2}
  \Big(adt - (r^2 + a^2)\frac{d\phi}{K}\Big)^2\Big].
\end{equation}
We mean $\ell$ is the AdS radius and the metric functions are defined as
\begin{subequations}
\begin{align}
f(r) &= (1 - \alpha^2r^2)\Big(1 - \frac{2m}{r} + \frac{a^2 + e^2 + g^2}{r^2}\Big)
 + \frac{r^2 + a^2}{\ell^2},\\
g(\theta)& = 1 + 2m\alpha\cos\theta + \Big(\alpha^2(a^2 + e^2 + g^2) - \frac{a^2}{\ell^2}\Big)\cos^2\theta,\\
& \Sigma = 1 + \frac{a^2}{r^2}\cos^2\theta,\quad
\Omega = 1 + \alpha r\cos\theta,
\end{align}
\end{subequations}
where $m$ is the black hole mass, $\alpha$ is a parameter which measures the acceleration of the black hole and $e$ and $g$ are the parameters denoting the electric and magnetic charges. 
We would like to consider the rotating accelerating black holes.
Then we leave the accelerating parameters $\alpha$ and the rotating parameter $a$ and set the others are zero $e = g = 0$. 
By redefining the scale, as 
$t/\ell\eqqcolon\tilde t$, $r/\ell\eqqcolon\tilde r$, $\alpha\ell\eqqcolon\tilde\alpha$, $m/\ell\eqqcolon\tilde m$ and $a/\ell\eqqcolon\tilde a$, we can eliminate the radius $\ell$ in the expression.
The rescaled metric $ds^2 = \ell^2d\tilde s^2$ is 
\begin{equation}
\label{eq:sec2_rescaledmetric}
d\tilde s^2
= \frac1{\Omega^2}\Big[-\frac{\tilde f(\tilde r)}{\Sigma}
 \Big(d\tilde t - \tilde a\sin^2\theta\frac{d\phi}{K}\Big)^2 + \frac{\Sigma}{\tilde f(\tilde r)}d\tilde r^2 + \frac{\Sigma}{g(\theta)}\tilde r^2d\theta^2
 + \frac{g(\theta)}{\Sigma}\frac{\sin^2\theta}{\tilde r^2}\Big(\tilde ad\tilde t - (\tilde r^2 + \tilde a^2)\frac{d\phi}{K}\Big)^2
\Big].
\end{equation}
In the following we write above rescaled quantities and functions without tilde $\tilde{\quad}$.

\subsection{Action}
In the metric \eqref{eq:sec2_rescaledmetric}, $\theta$ and $\varphi$ are coordinates on the sphere $S^2$. 
We would like to consider a prove string moving on the equator of this sphere at constant angular momentum.

\paragraph{World-sheet coordinates}
The motion of the string is restricted on $\theta = \pi/2$.
For $\theta = \pi/2$, the metric becomes,
($\Omega = \Sigma = g(\pi/2) = 1$)
\begin{equation}
\label{eq:rescaledmetric_on_equator}
ds^2_\mathrm{ind}
= -f(r)\Big(dt - a\frac{d\phi}{K}\Big)^2 
 + \frac{dr^2}{f(r)}
 + \frac{1}{r^2}\Big(adt - (r^2 + a^2)\frac{d\phi}{K}\Big)^2.
\end{equation}
Here let us consider the behavior in the infinity (AdS boundary).
At $r\rightarrow\infty$ and $\alpha = 0$, the first and the last terms of \eqref{eq:rescaledmetric_on_equator} are
\begin{equation}
-r^2\Big(dt - a\frac{d\phi}{K}\Big)^2
= -r^2\Big(\Big(1 + \frac{a}{K}\tilde V\Big)dt - a\frac{d\Phi}{K}\Big)^2,\quad
\frac1{r^2}\Big(-r^2\frac{d\phi}{K}\Big)^2
= r^2\Big(\frac{\tilde V}{K}dt - \frac{d\Phi}{K}\Big)^2
\end{equation}
where a new coordinate is defined as $\Phi\coloneqq\phi+ \tilde Vt$.
Since the cross terms
\begin{equation}
2r^2\frac{a}{K}\Big(1 + \frac{a}{K}\tilde V\Big) - 2r^2\frac{\tilde V}{K^2}
= \frac{2r^2}{K^2}(aK + (a^2 - 1)\tilde V)
\end{equation}
must be canceled. 
Then we set $\displaystyle \tilde V = \frac{aK}{1-a^2}$.
In these coordinates we parametrize the worldsheet as
\begin{equation}
t = \tau,\quad
r = \sigma,\quad
\Phi = V\tau + \xi(\sigma);\quad
\phi = \omega\tau + \xi(\sigma),
\end{equation}
where in the last expression we defined the string angular momentum as 
$\omega\coloneqq V - \tilde V$.

By this parametrization, the induced metric on string world sheet $(\tau,\sigma)$ is
\begin{align}
ds_\mathrm{ind}^2
&= \bigg[-f(\sigma)\Big(1 - \frac{a\omega}{K}\Big)^2
 + \frac{1}{\sigma^2}\Big(a - (\sigma^2 + a^2)\frac{\omega}{K}\Big)^2\bigg]d\tau^2 \nonumber\\
&\qquad
 + \bigg[\frac1{f(\sigma)}
 + \Big(-f(\sigma)a^2 + \frac{(\sigma^2 + a^2)^2}{\sigma^2}\Big)\frac{\xi'^2}{K^2}\bigg]d\sigma^2
 \nonumber\\
&\qquad
 + 2\bigg[f(\sigma)\Big(1 - \frac{a\omega}{K}\Big)a - \frac{\sigma^2 + a^2}{\sigma^2}\Big(a - (\sigma^2 + a^2)\frac{\omega}{K}\Big)\bigg]\frac{\xi'}{K}d\tau d\sigma.
\end{align}
By calculating the determinant 
\begin{align}
-\det(g_\mathrm{ind})
&= \Big[f\Big(1 - \frac{a\omega}{K}\Big)^2
   - \frac{1}{\sigma^2}\Big(a - (\sigma^2 + a^2)\frac{\omega}{K}\Big)^2\Big]
  \Big[\frac{1}{f} - \Big(fa^2 - \frac{(\sigma^2 + a^2)^2}{\sigma^2}\Big)\frac{\xi'^2}{K^2}\Big]\nonumber\\
&\qquad
 + \Big[f\Big(1 - \frac{a\omega}{K}\Big)a - \frac{\sigma^2 + a^2}{\sigma^2}\Big(a - (\sigma^2 + a^2)\frac{\omega}{K}\Big)\Big]^2\frac{\xi'^2}{K^2}\nonumber\\
&= \Big(1 - \frac{a\omega}{K}\Big)^2
 - \frac{1}{\sigma^2f}\Big(a - (\sigma^2 + a^2)\frac{\omega}{K}\Big)^2
 + \sigma^2f\frac{\xi'^2}{K^2},
\end{align}
we obtain the Nambu-Goto Lagrangian as
\begin{equation}
\mathcal L\coloneqq\sqrt{-\det(g_\mathrm{ind})}
= \sqrt{\Big(1 - \frac{a\omega}{K}\Big)^2
 - \frac{1}{F(\sigma)}\Big(a - (\sigma^2 + a^2)\frac{\omega}{K}\Big)^2
 + F(\sigma)\frac{\xi'^2}{K^2}}.
\end{equation}
Here we defined $F(\sigma) \coloneqq \sigma^2f(\sigma)$.

Since
$\displaystyle\frac{\partial\mathcal L}{\partial\xi} = 0$,
the equation of motion can be solved easily as
\begin{equation}\label{eq:eom_to_xi}
\frac{d}{d\sigma}\Big(\frac{\sigma^2f}{K^2}\frac{\xi'}{\mathcal L}\Big) = 0 \quad\implies\quad
\frac{\sigma^2f}{K^2}\frac{\xi'}{\mathcal L} = c \quad (c\in\mathbb R_{<0}).
\end{equation}
where the sign of $c$ is negative because we now focus on the inner of the black hole horizon where $\sigma^2f(\sigma) < 0$ and $\xi'$ is positive.
This can be solved for $\xi'$; 
\begin{equation}
\xi'(\sigma) 
= \frac{cK^2}{F}\sqrt\frac{(a - (\sigma^2 + a^2)\omega/K)^2 - F(1 - a\omega/K)^2}{c^2K^2 - F}.
\end{equation}

Let $\sigma_0$ be the solution of
$(a - (\sigma_0^2 + a^2)\omega/K)^2 - F(\sigma_0)\cdot(1 - a\omega/K)^2 = 0$.
(a point where the numerator of the inner of the square root becomes zero.)
We write $F_0 \coloneqq F(\sigma_0)$. 
Since the denominator must be zero at the same point $\sigma = \sigma_0$, $c$ is determined by the value of $\sigma_0$ as 
\begin{equation}
c^2K^2 = F_0
= \frac{(a - (\sigma_0^2 + a^2)\omega/K)^2}{(1-a\omega/K)^2}.
\end{equation}
Then, by eliminating the constant $c$ from the expression, 
\begin{equation}
\xi' = \frac{F_0}{F}\sqrt\frac{(a - (\sigma^2 + a^2)\omega/K)^2 - F(1 - a\omega/K)^2}{F_0 - F}.
\end{equation}
By the relation \eqref{eq:eom_to_xi}, the Lagrangian is 
\begin{equation}
\mathcal L
= \sqrt\frac{(a - (\sigma^2 + a^2)\omega/K)^2 - F(1 - a\omega/K)^2}{F_0 - F}.
\end{equation}
By defining $\tilde\omega\coloneqq\omega/K$, we find the deficit parameter $K$ effects only through the rescaled angular momentum $\tilde\omega$,
\begin{equation}
\mathcal L
= \sqrt\frac{(a - (\sigma^2+a^2)\tilde\omega)^2 - F(1-a\tilde\omega)^2}{F_0 - F}.
\end{equation}

The growth of the Nambu-Goto action is 
\begin{equation}
\frac{dS_\mathrm{NG}}{dt}(\omega,m,a,\alpha)
= T_\mathrm{str}\int_{r_{-}}^{r_{+}}d\sigma
  \sqrt\frac{(a - (\sigma^2+a^2)\tilde\omega)^2 - F(1-a\tilde\omega)^2}{F_0 - F},
\end{equation}
where $r_\pm$ is the two positive solutions of $F(r) = 0$.

\begin{example}
In non-rotating limit $a = \alpha = 0$ and $K=1$, this surely gives the Lagrangian for Schwartzchild black holes:
\begin{equation}
\mathcal L\big|_{a=0}
= \sqrt{1 - \frac{\sigma^2}{f}\omega^2 + \sigma^2f\xi'^2},\qquad
\xi' = \frac{cK^2}{F}\sqrt\frac{\sigma^4\omega^2 - F}{c^2 - F},\quad
F(\sigma) = \sigma^2f(\sigma) = \sigma^4 + \sigma^2 - 2m\sigma.
\end{equation}
Its action growth is
\begin{equation}
\frac{dS_\mathrm{NG}}{dt}
= T_\mathrm{s}\int_{0}^{r_\mathrm{h}}d\sigma
  \sqrt\frac{\sigma^4\tilde\omega^2 - F}{\sigma_0^4\tilde\omega^2 - F},\quad
F(\sigma) = \sigma^4 + \sigma^2 - 2m\sigma.
\end{equation}
It corresponds to the case we considered in \cite{Nagasaki:2017kqe}. 
\end{example}

\section{Calculating the action}
\begin{figure}
	\begin{minipage}[h]{0.5\linewidth}
	\begin{center}
	\includegraphics[width=9.5cm]{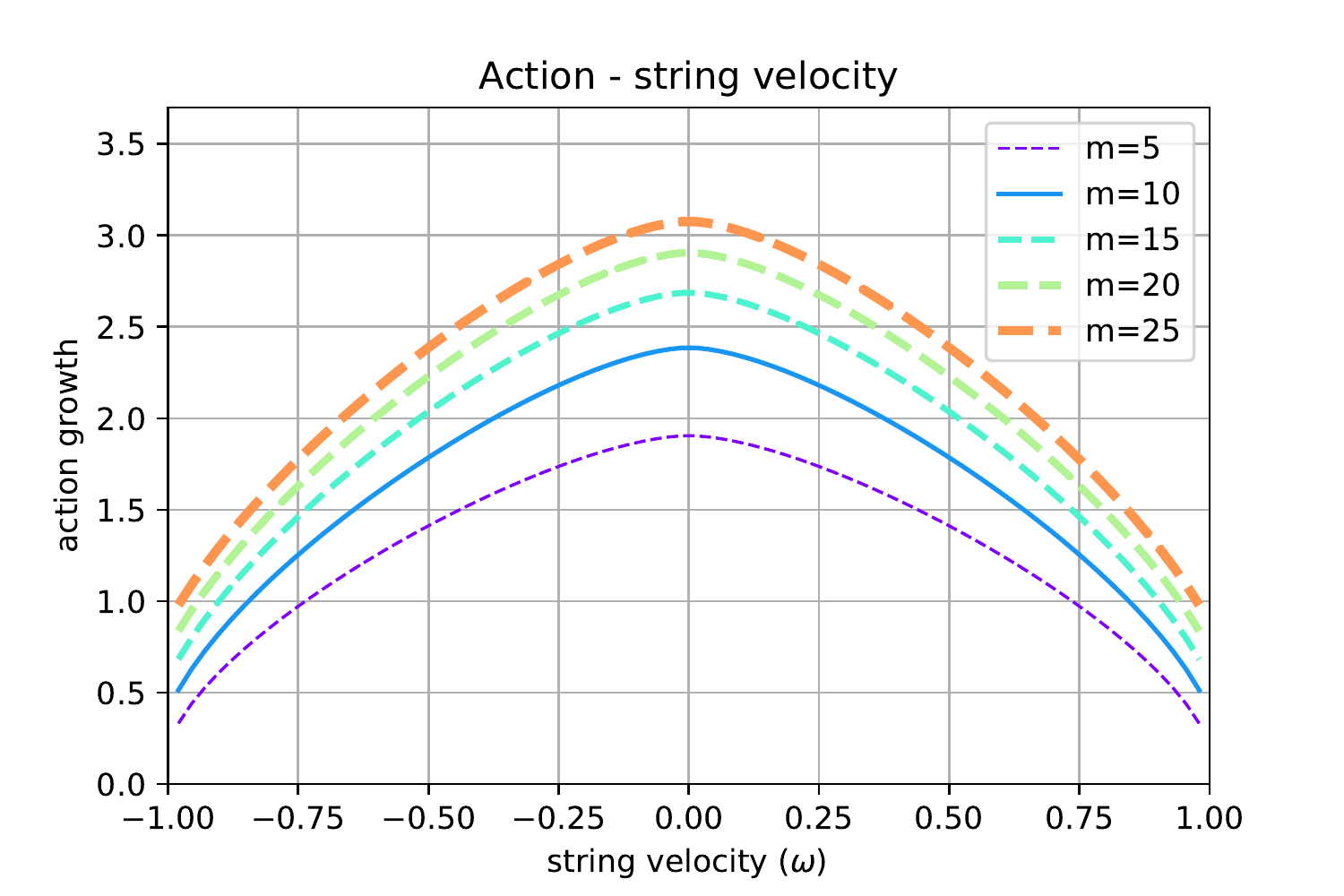}
	\caption{$a = 0, \alpha = 0.2$, different messes}
	\label{fig:cmetric_NGaction_omega_mvar_a0_alpha2}
	\end{center}
	\end{minipage}
\hspace{0\linewidth}
	\begin{minipage}[h]{0.5\linewidth}
	\begin{center}
	\includegraphics[width=9.5cm]{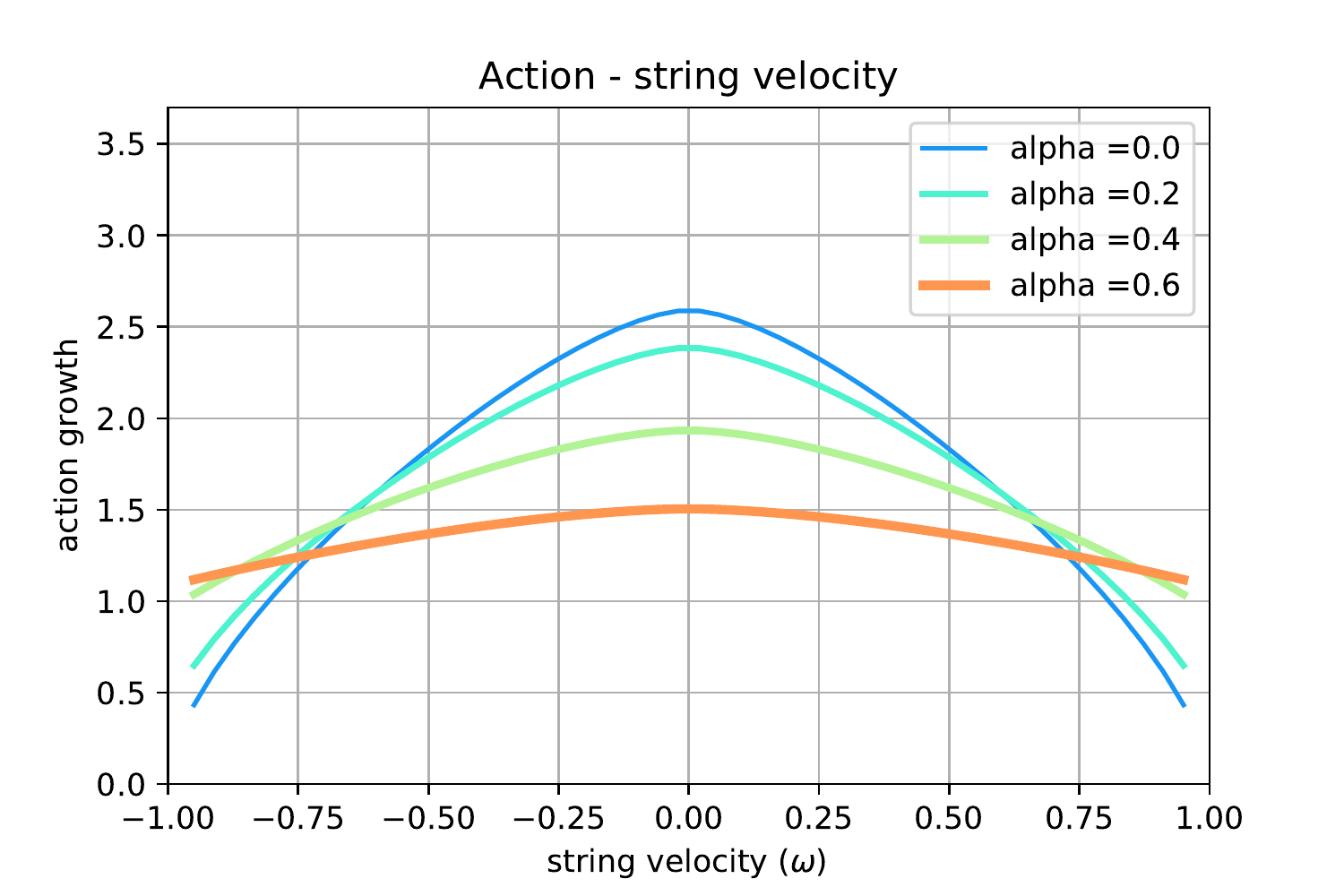}
	\caption{$m = 10, a = 0$, different acceleration}
	\label{fig:cmetric_NGaction_stringvelocity_alphavar_a0_mass10}
	\end{center}
	\end{minipage}
\end{figure}

\begin{figure}
\hspace{0.05\linewidth}
	\begin{minipage}[h]{0.45\linewidth}
	\begin{center}
	\includegraphics[width=9.5cm]{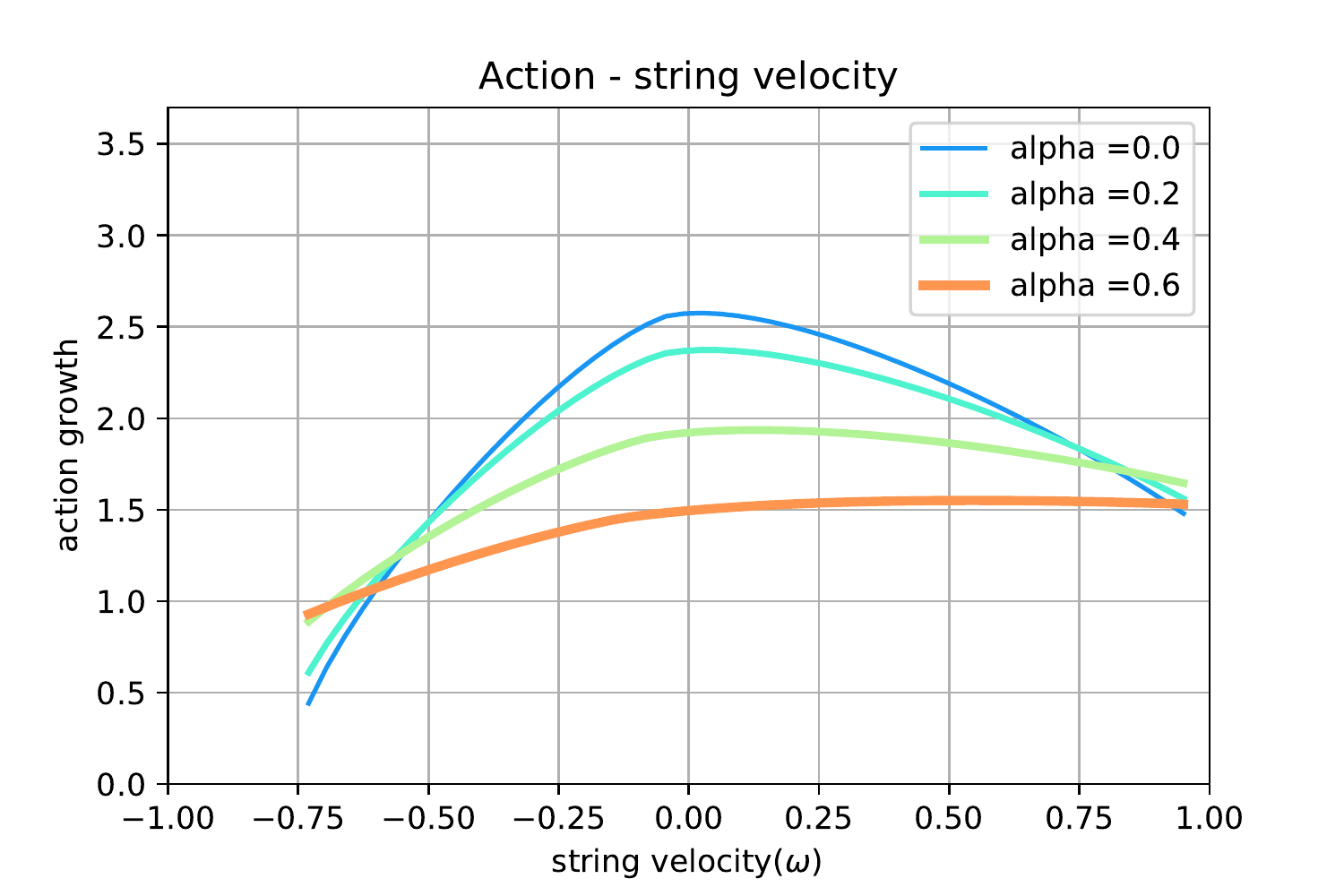}
	\caption{$m = 10, a = -0.3$, different acceleration}
	\label{fig:cmetric_NGaction_stringvelocity_alphavar_aneg3_mass10}
	\end{center}
	\end{minipage}
\end{figure}

The results of the numerical calculation are shown in the following figures
(Fig.\ref{fig:cmetric_NGaction_omega_mvar_a0_alpha0} - Fig.\ref{fig:cmetric_NGaction_stringvelocity_alphavar_aneg3_mass10}).

\paragraph{Schwarzschild black holes}
The first one (Fig.\ref{fig:cmetric_NGaction_omega_mvar_a0_alpha0}) is the case where $a = \alpha = 0$. 
This is the Schwarzschild black hole case. 
\paragraph{Kerr black holes}
As we can see in Fig.\ref{fig:cmetric_NGaction_omega_mvar_a2_alpha0},
the rotation of the black hole causes the effect of the distortion of the distribution. 
Especially, it does not affect the maximum value of the distribution by comparing with Fig.\ref{fig:cmetric_NGaction_omega_mvar_a0_alpha0} but the rotation causes the ``shift" to {\bf the opposite direction to the black hole angular momentum} except the maximum point of the graph.

\paragraph{Accelerating black holes}
The acceleration of the black hole makes the string effect smaller.
Especially, it decreases the maximum value of it by comparing Fig.\ref{fig:cmetric_NGaction_omega_mvar_a0_alpha2} (no acceleration) and Fig.\ref{fig:cmetric_NGaction_omega_mvar_a0_alpha0} (nonzero acceleration).
The rate of this reduction is shown in Fig.\ref{fig:cmetric_NGaction_stringvelocity_alphavar_a0_mass10} in no rotating cases  
and Fig.\ref{fig:cmetric_NGaction_stringvelocity_alphavar_aneg3_mass10} for rotating cases.
As mentioned in the Kerr black hole case, the graph is shifted to {\bf the opposite to the black hole rotation}.

\begin{figure}[h]
	\begin{minipage}[h]{0.5\linewidth}
	\begin{center}
	\includegraphics[width=9.5cm]{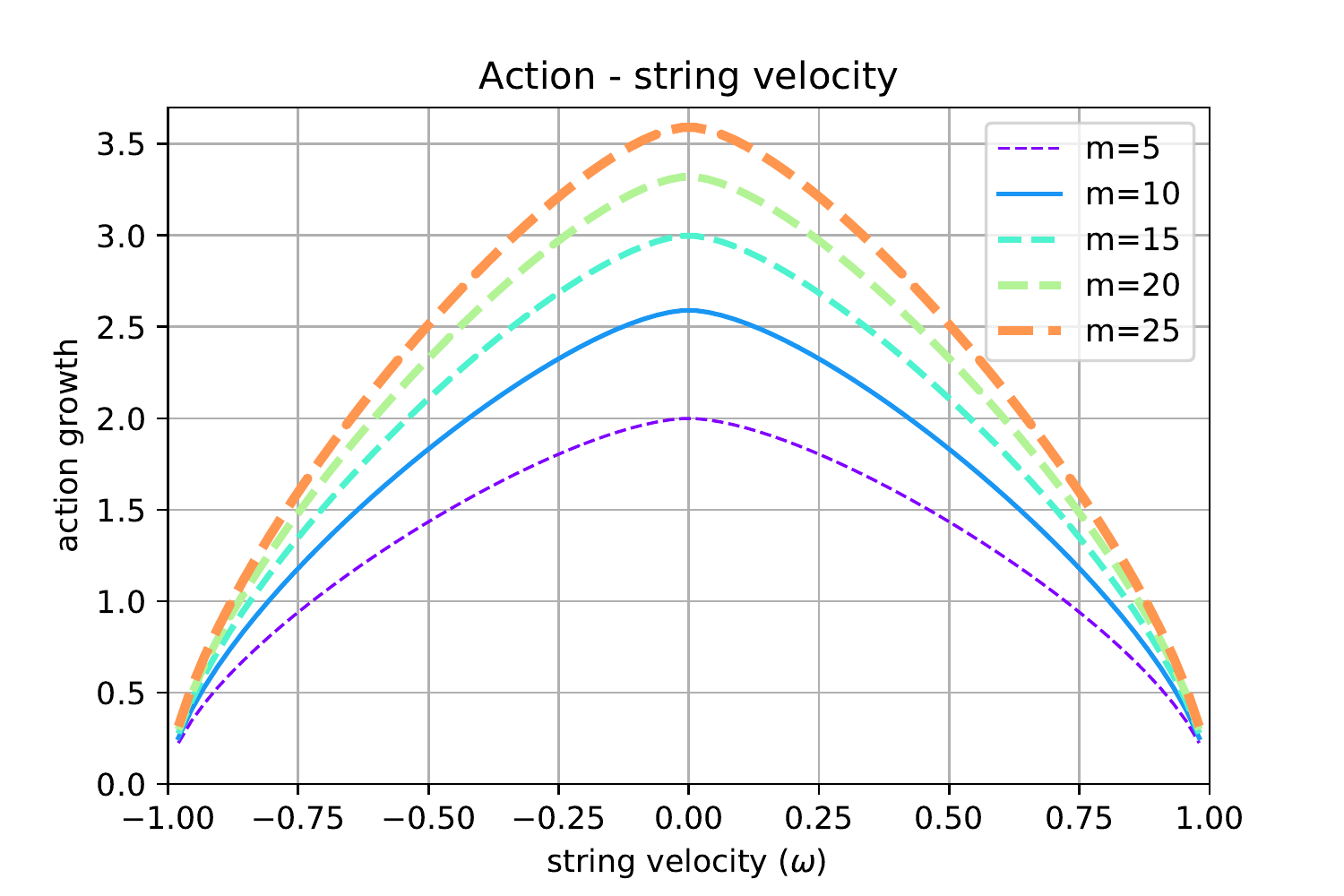}
	\caption{$a = \alpha = 0$ and different masses}
	\label{fig:cmetric_NGaction_omega_mvar_a0_alpha0}
	\end{center}
	\end{minipage}
\hspace{0\linewidth}
	\begin{minipage}[h]{0.5\linewidth}
	\begin{center}
	\includegraphics[width=9.5cm]{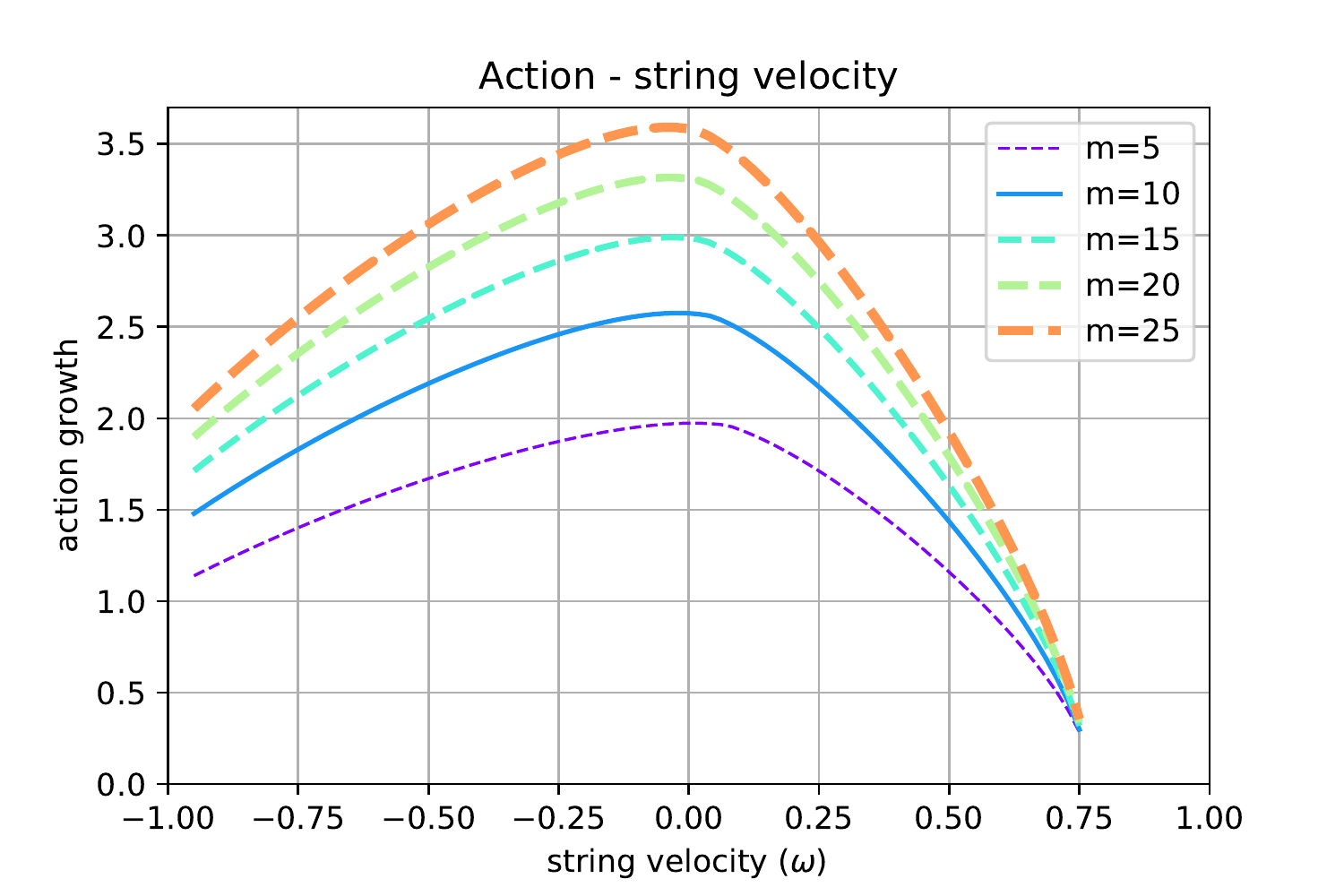}
	\caption{$a = 0.2, \alpha = 0$, different messes}
	\label{fig:cmetric_NGaction_omega_mvar_a2_alpha0}
	\end{center}
	\end{minipage}
\end{figure}

\newpage
\section{Extremum point of the distribution}
Since the integral region $r_\pm$ is independent of the value $\omega$, the partial derivative is
\begin{equation}
\frac{\partial}{\partial\omega}
\Big(T_\mathrm{str}^{-1}\frac{dS_\mathrm{NG}}{dt}(\omega)\Big)
= \frac12\int_{r_{-}}^{r_{-}}d\sigma L^{-1/2}\frac{\partial L}{\partial\omega}.
\end{equation}
Then in order to find the extremum point we have to find the point where 
$\displaystyle \frac{\partial L}{\partial\omega} = 0$ is satisfied.

\paragraph{By fractorized form}
We would like to find the above point by the method of Lagrange multipliers.
We use the variables $x\coloneqq \omega$, $y\coloneqq \sigma_0$.
\begin{equation}
L(x,y) \coloneqq
 (1-ax)^2 + xs_1(y)\frac{2a(1-ax) - x(y^2 + \sigma^2)}{f_1 + f_2s_1(y) + f_3s_2(y) + f_4s_3(y)},
\end{equation} 
where $f_n$ are constant (independent of $\sigma$) determined by the properties of black holes (mass, angular momentum and acceleration) and $s_n$ are functions of $y$,
\begin{equation}
s_n(y) = \sum_{k=0}^{n}\sigma^{n-k}y^k.
\end{equation}
We would like to find the maximal point with the constraint
\begin{equation}
\tilde F(x,y)\coloneqq F(y)(1-ax)^2 - (a - (y^2 + a^2)x)^2 = 0.
\end{equation}
We define the function 
\begin{equation}
\tilde L(x,y,\lambda)\coloneqq L(x,y) + \lambda\tilde F(x,y)
\end{equation}
and would like to find the extremum of this function.

\begin{example}[Schwarzchild case]
For $a=0$ and $\alpha = 0$, the problem is simplified as 
\begin{equation}
L(x,y,\lambda) = 1 - x^2s_1\frac{y^2 + \sigma^2}{f_1 + s_1 + s_3},\quad
\tilde F(x,y) = F_\mathrm{sch}(y) - y^3x^2 = 0,\quad
F_\mathrm{sch}(y) = y^3 + y - 2m.
\end{equation} 
The differentials are
\begin{align}
\frac{\partial\tilde L}{\partial x}
&= -2x\Big(s_1\frac{(y^2 + \sigma^2)}{s_3 + s_1 - 2m} + \lambda y^3\Big),\\
\frac{\partial\tilde L}{\partial y}
&= -x^2\frac{(s_1'(y^2 + \sigma^2) + 2ys_1)(s_3 + s_1 - 2m) - s_1(y^2 + \sigma^2)(s_3' + s_1')}{(s_3 + s_1 - 2m)^2} + \lambda(3(1-x^2)y^2 + 1),\\
\frac{\partial\tilde L}{\partial\lambda}
&= (1-x^2)y^3 + y - 2m.
\end{align}
Then $x = \lambda = 0$ is a solution and the last equation has the unique positive solution for $y$ when $x=0$.
It is the same value to the horizon.
\qed
\end{example}

In general, 
\begin{align}
\tilde L(x,y,\lambda) 
&= (1-ax)^2 + xs_1\frac{2a(1-ax) - x(y^2 + \sigma^2)}{f_1 + f_2s_1 + f_3s_2 + f_4s_3}
 + \lambda\Big((a - (y^2 + a^2)x)^2 - (1-ax)^2\sum_{n=0}^{4}f_ny^n\Big).
\end{align}
The derivatives are
\begin{align}
\frac{\partial\tilde L}{\partial x}
&= -2a(1-ax) + s_1\frac{2a(1-2ax) - 2x(y^2 + \sigma^2)}{f_1 + f_2s_1 + f_3s_2 + f_4s_3}
 + 2\lambda\Big(-(y^2 + a^2)(a - (y^2 + a^2)x) + a(1-ax)\sum_{n=0}^{4}f_ny^n\Big),\nonumber\\
\frac{\partial\tilde L}{\partial y}
&= x\frac{s_1'(2a(1-ax) - x(y^2 + \sigma^2)) - 2xys_1}{f_1 + f_2s_1 + f_3s_2 + f_4s_3}
 - xs_1(2a(1-ax) - x(y^2 + \sigma^2))\frac{\sum_{n=0}^{4}f_{n+1}s_n'}{(\sum_{n=0}^{4}f_{n+1}s_n)^2}\nonumber\\
&\hspace{8cm}
 + \lambda\Big(-4xy(a- (y^2 + a^2)x) - (1-ax)\sum_{n=1}^{4}nf_ny^{n-1}\Big),\nonumber\\
\frac{\partial\tilde L}{\partial\lambda}
&= -[(1-ax)^2(f_0 - a^2) + (1-ax)^2f_1y + (f_2 + 2ax)(1-ax)y^2 + (1-ax)^2f_3y^3 + ((1-ax)^2f_4 - x^2)y^4].
\end{align}

We would like to confirm that the maximal point is independent of $\alpha$.

\begin{align}
\frac{\partial\tilde L}{\partial x}
&= -2xs_1\frac{(y^2 + \sigma^2)}{f_1 + s_1 + f_3s_2 + f_4s_3}
 + 2\lambda y^4x,\\
\frac{\partial\tilde L}{\partial y}
&= x\frac{s_1'(-x(y^2 + \sigma^2)) - 2xys_1}{f_1 + s_1 + f_3s_2 + f_4s_3}
 - xs_1(-x(y^2 + \sigma^2))\frac{\sum_{n=0}^{4}f_{n+1}s_n'}{(\sum_{n=0}^{4}f_{n+1}s_n)^2}
 + \lambda\Big(4x^2y^3 - \sum_{n=1}^{4}nf_ny^{n-1}\Big),\\
\frac{\partial\tilde L}{\partial\lambda}
&= x^2y^4 - \sum_{n=0}^{4}f_ny^n.
\end{align}
The first and the second equations have solution $x = \lambda= 0$.
In this case, the third equation gives $y$ equal to the horizon value.
We conclude that for non-rotating black holes $a=0$ the acceleration does not effect the maximal point.

\section{Discussion}
In this work we studied the effects of the probe string on the accelerating black hole spacetime, especially it has the angular momentum.
We summarize the main results.
The basic behavior of holographic complexity on the probe string is studied in \cite{Nagasaki:2017kqe}: 
\begin{enumerate}
\item[(0)] The effect of the string on the growth of complexity is minimized when the probe string is stationary.
\end{enumerate}
The new properties we found are:
\begin{enumerate}
\item[(1)] {\bf Effect of the angular momentum}\\
The rotation of the black hole causes the asymmetric behavior in the direction of string motion. 
In the past work \cite{Nagasaki:2018csh} we only know that the black hole angular momentum causes the shift of the distribution between the growth of the action and the string motion.
Now, in addition to this, we found the maximal value is independent of the black hole angular momentum (Fig.\ref{fig:cmetric_NGaction_omega_mvar_a0_alpha0} and Fig.\ref{fig:cmetric_NGaction_omega_mvar_a2_alpha0}) by the numerical methods.
Further, it will shed light on the relation between the rotation and the black hole acceleration as stated below. 

\item[(2)] {\bf Effect of the acceleration}\\
The acceleration of the black hole makes the string effect smaller.
We also studied the relation between the effect of the black hole angular momentum and that of the acceleration by the analytic method.
We found that when the black hole does not have the angular momentum the acceleration $\alpha$ does not change the maximal point (with respect to the string motion $\omega$).
This result can also be checked visually by comparing Fig.\ref{fig:cmetric_NGaction_omega_mvar_a2_alpha0} ($a\neq0$) and Figs.\ref{fig:cmetric_NGaction_omega_mvar_a0_alpha2}, \ref{fig:cmetric_NGaction_stringvelocity_alphavar_a0_mass10} ($a=0$).
\end{enumerate}

In addition to these, we studied the maximal of the Nambu-Goto action in analytic way.
We confirmed this results found in \cite{Nagasaki:2017kqe} analytically and confirmed that Property (0) is surely satisfied.
We also found that the accelerating decreases the effect of the probe string while it does not shift the maximal point.
This result is also confirmed numerically in Fig.\ref{fig:cmetric_NGaction_stringvelocity_alphavar_a0_mass10}.

\section*{Acknnowledgments}
This research is supported by Department of Physics, Toho University.


\providecommand{\href}[2]{#2}\begingroup\raggedright\endgroup

\end{document}